# BLAST Autonomous Daytime Star Cameras

Marie Rex*[a], Edward Chapin[b], Mark J. Devlin[a], Joshua Gundersen[c], Jeff Klein[a], Enzo Pascale[d], Donald Wiebe[d]

[a] University of Pennsylvania, Department of Physics and Astronomy, 209 S. 33rd St., Philadelphia, PA 19104
[b] University of British Columbia, Department of Physics and Astronomy, 6224 Agriculture Rd., Vancouver, BC V6T-1Z1, Canada
[c] University of Miami, Department of Physics, 1320 Campo Sano Drive, Coral Gables, FL 33146
[d] University of Toronto, Department of Physics and Astronomy, 60 St. George St., Toronto, ON M5S-1A7, Canada

**ABSTRACT**

We have developed two redundant daytime star cameras to provide the fine pointing solution for the balloon-borne submillimeter telescope, BLAST. The cameras are capable of providing a reconstructed pointing solution with an absolute accuracy < 5″. They are sensitive to stars down to magnitudes ~ 9 in daytime float conditions. Each camera combines a 1 megapixel CCD with a 200 mm f/2 lens to image a 2º × 2.5º field of the sky. The instruments are autonomous. An internal computer controls the temperature, adjusts the focus, and determines a real-time pointing solution at 1 Hz. The mechanical details and flight performance of these instruments are presented.

## 1. INTRODUCTION

We have developed two redundant daytime star cameras to provide the fine pointing solution for the "Balloon-borne Large-Aperture Submillimeter Telescope" (BLAST)[1]. The BLAST gondola consists of an outer frame, which is pointed in azimuth using a flywheel/pivot system, and an inner frame which moves independently in elevation (Fig. 1). The inner frame houses the telescope, detectors, and the two star cameras, and is pointed relative to the outer frame using servo motors mounted inside the elevation bearing. The instrument combines a 2 m telescope with bolometer detector arrays at 250, 350, and 500 microns with resolutions of 30″, 41″, and 59″ respectively. Each array has a 7′ × 13′ field of view, requiring maximum in-flight absolute pointing errors of < 2′. Post-flight pointing requirements are more demanding. An rms reconstructed pointing accuracy of 8″ is necessary in order to successfully extract point source fluxes from the final maps[2]. To achieve these requirements, BLAST uses an array of pointing sensors including GPS, gyroscopes, a digital sun sensor, a magnetometer, and two star cameras.

The primary absolute pointing sensors for BLAST are the two integrating star cameras that are mounted above the telescope on the inner frame. The cameras image a 2º × 2.5º field of the sky, and are sensitive to stars down to magnitude ~ 9 in typical daytime float conditions. They are capable of providing a reconstructed pointing solution with an accuracy < 5″. The instruments are autonomous. An internal computer controls the temperature, adjusts the focus, and determines a real-time pointing solution at 1Hz.

At typical balloon altitudes of ~ 40 km there is still enough atmosphere to create a bright daytime sky background at optical wavelengths. The challenge is to design a system that can detect enough stars in a given field of view to determine a pointing solution, have a short enough integration time to accommodate the scanning modes of the telescope, and have a CCD with sufficient sensitivity, well depth, and dynamic range to detect stars above the background atmosphere in the required integration time. Furthermore, temperature fluctuations can change the lens focus, requiring a mechanism for adjusting the focal element in flight.

*contact author at madamson@student.physics.upenn.edu

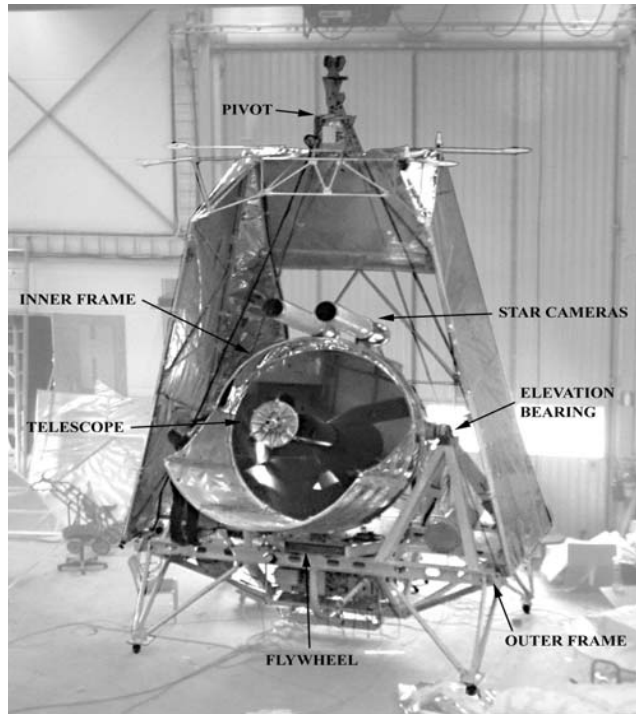

Fig. 1: A picture of the BLAST payload. The gondola consists of an outer frame which is pointed in azimuth using a flywheel/pivot system, and an inner frame which moves independently in elevation. The inner frame houses the telescope, detectors, and the two star cameras, and is pointed relative to the outer frame using servo motors mounted inside the elevation bearing.

The BLAST payload had a successful four day flight from Kiruna, Sweden in June 2005. Both star cameras worked reliably throughout the flight. The star camera pointing solution was used to reconstruct the gondola attitude to 5″ accuracy. We present the details of the BLAST star camera design and performance in this paper.

## 2. SYSTEM DESCRIPTION

Each star camera system is an assembly of many components selected to satisfy the pointing requirements and provide reliable operation in flight. An internal computer runs sophisticated software enabling the unit to run autonomously. It calculates pointing solutions in real time and communicates them to the flight computer via Ethernet. This section provides a description of the components in each star camera unit. A mechanical drawing of the assembly is shown in Figure 2.

### 2.1 CCD camera

Though both star cameras are nearly identical, they use different CCD cameras. The first unit (ISC) uses the QImaging PMI 1401and the second unit (OSC) uses the QImaging Retiga EXi. The specifications of these two CCD cameras are listed in Table 1. The PMI 1401 has a deeper pixel well. It saturates at 45,000 $e^-$ while the Retiga saturates at only 18,000 $e^-$. This enables the ISC to integrate longer before saturation, and therefore detect stars in brighter sky conditions. However, the Retiga Exi has more bits/$e^-$, therefore the OSC is more sensitive in dimmer conditions. Both cameras are high resolution, with $10^6$ pixels measuring ~ 7 μm × 7 μm. Combined with the lens optics, the small pixel size facilitates a precise pointing solution, and reduces the background signal due to sky brightness in individual pixels. Both CCDs have a peak quantum efficiency of ~ 65 % at 600 nm, with maximum spectral response from 400 nm – 850 nm.

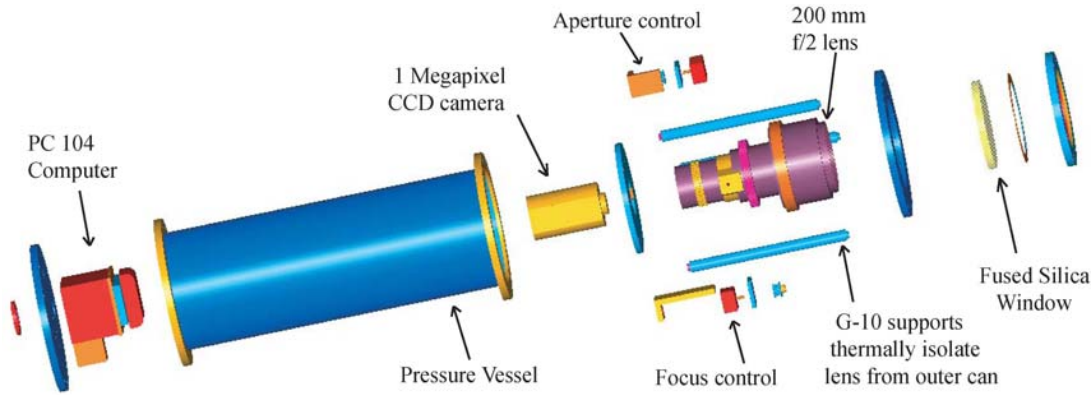

Fig. 2: Mechanical drawing of star camera assembly. The design is based on an integrating CCD camera coupled to 200 mm f/2 lens with a 2º × 2.5º field of view. The entire system in contained in a pressure vessel.

### 2.2 Optics

BLAST is a mapping experiment. The telescope scans sources with an azimuthal slew rate of 10′/sec to optimize detector response. This slew rate puts an upper limit on the star camera exposure time. A 100 mm Nikon lens with a 200 mm focal length was chosen to reduce the necessary exposure time. When coupled with the ~ 7 mm × 7 mm CCD array, the star camera has a 2º × 2.5º field of view and a 7″ pixel scale.

The lens has a Nikon R60 filter which filters out light at less than 600 nm. This red filter greatly reduces the background level due to sky brightness, but also rejects some of the light coming from stars. The Air Force software package MODTRAN has been used by many groups developing star cameras to simulate expected sky background levels at balloon altitudes. Comparing these predictions to star count models indicates this filter reduces background sky brightness more than the typical star brightness, improving the signal to noise of star detections[3].

The background radiation is further attenuated by a 4 ft baffle attached to the front of each camera (Fig. 3). The baffle permits the star camera to acquire pointing solutions as close as 60º from the sun. It uses knife-edged aluminum disks to reject scattered light. The baffle length prevents all sources > 7º from the optical axis from directly illuminating the window. The internal geometry eliminates primary reflections from sources > 10º from the optical axis. The disks are sand-blasted and anodized to minimize reflections inside the baffle. The structure is held together by G-10 tube to provide lightweight support.

Table 1: Specifications of the two QImaging CCD cameras used in the BLAST star cameras. The ISC incorporates the QImaging PMI 1401, while the OSC uses the more recent model, Retiga EXi.

|  | ISC | OSC |
| --- | --- | --- |
| **QImaging Camera** | PMI 1401 | Retiga EXi |
| **CCD Sensor** | Kodak KAF-1401 | Sony ICX285 |
| **Light Sensitive Pixels** | 1312 × 1024 | 1360 × 1036 |
| **Pixel Size** | 6.8 μm × 6.8 μm | 6.45 μm × 6.45 μm |
| **Digital Output** | 14 bit | 12 bit |
| **Pixel Well Depth** | 45,000 $e^-$ | 18,000 $e^-$ |
| **Read Noise** | 20 $e^-$ | 8 $e^-$ |

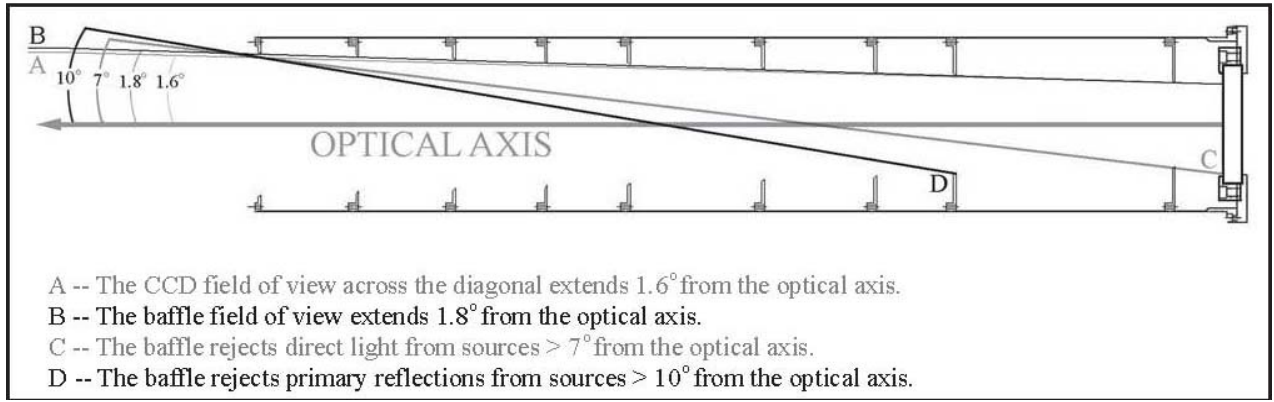

Fig. 3: Cross section of the star camera baffle. The baffle uses thin aluminum disks to reject scattered light. The baffle length prevents all sources > 7º from the optical axis from directly illuminating the window. The internal geometry eliminates primary reflections from sources > 10º from the optical axis.

The signal to noise is also improved by minimizing the number of pixels contained in a star image. This is done by keeping the lens in focus throughout the flight. Dry ice tests revealed that the lens changes focus as a function of temperature. Due to large temperature fluctuations during ascent, the focus is adjustable in flight via stepper motor. The lens is thermally isolated from the external assembly with G-10 posts. Each unit also contains a thermostat and heater to regulate the temperature of the lens and prevent the focal element from freezing.

**2.3 Electronics**

A PC-104 computer is mounted inside each star camera pressure vessel. The computer runs software that processes the CCD images, commands the stepper motors, and regulates the lens temperature. The system reports the right ascension and declination of the center of the image to the flight computer over an Ethernet line. It also outputs video images which are sent down by transmitters on the gondola, providing a visual confirmation of the star camera field. A schematic of the electronic subsystems is shown in Figure 4.

The camera can operate in two trigger modes. An internal trigger is useful for testing the star camera in the lab, independent from the rest of the payload. However, during flight, accurate timing of each frame is required. Therefore, when it is integrated with the gondola, it uses an external trigger which is generated by the main flight computer.

The temperature and pressure of the unit are monitored using a programmable USB based DAQ module. The module has 12–bit analog input channels that read out the temperature and pressure. Its digital output is programmed to enable the heater when the lens drops below 10º Celsius.

Two 0.9º stepper motors are used to adjust the lens focus and aperture. The motors are controlled by AllMotion EZ stepper drivers which are commanded by the software through the RS232 serial line. The unit is launched with the aperture fully open, and with the lens focused at room temperature infinity. An optical limit switch indicates this "home position" for reference during the flight, but an auto-focusing routine is run once the payload reaches float altitude. This routine measures the flux of the brightest source in the field of view for 100 steps around the home position and chooses the position that maximizes this flux.

The computer displays each image on the monitor and indicates the results of the analysis. Sources are identified in boxes, and labeled with their visual magnitude. The RA and DEC solution is displayed and converted to telescope coordinates. Star camera temperature, pressure and heater status are also displayed for reference. This image is converted to video and transmitted down during flight for troubleshooting.

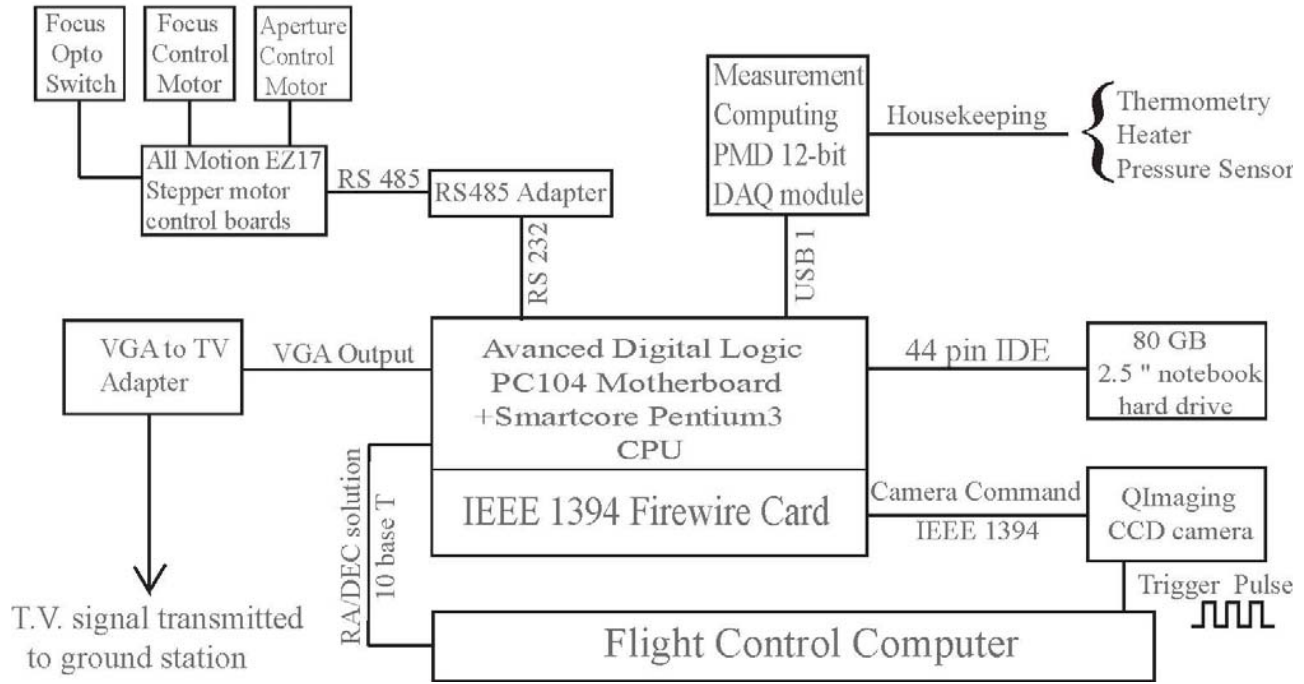

Fig. 4: Block diagram of the electronic subsystems. The star camera computer controls the camera and stepper motors, regulates the temperature, and outputs a video signal for troubleshooting from the ground. The star camera receives a square wave trigger pulse from the flight computer, and returns a corresponding pointing solution via Ethernet line.

**2.4 Software**
A pattern recognition algorithm runs on board to calculate the absolute pointing of the star camera in real-time. When each CCD frame is read in, any baseline line gradient across the image must be subtracted. All previously identified bad pixels are set to the mean value. Sources are then selected based on a threshold signal to noise. The centroid positions of the sources are used to calculate relative angular offsets between sources based on the measured angular pixel scale. These angular offsets are compared to a star catalog to find the best match. Initially the camera is given an approximate pointing solution from another pointing sensor. This technique allows the software to search only a small portion of the star catalog. After the initial match, the software uses previous star camera solutions for reference.

The software requires at least three sources for a unique in-flight solution. After the flight, however, stars in single and double source frames can be matched to previous frames to provide pointing solutions. To maintain > 3 sources in a frame the star camera can require an integration time of up to 300 ms in bright daytime float conditions. Given the scan rate of the gondola, this exposure time causes sources to be smeared across the camera during scans, and in-flight solutions are unreliable. Therefore, star camera triggers are timed to obtain pointing solutions when the telescope changes directions at the ends of each scan. Between solutions, the relative pointing is interpolated using the gyroscopes. Shorter integration times are used when the gondola is in motion, where images can provide post-flight reconstruction of the telescope orientation.

## PERFORMANCE SUMMARY

The ISC has flown as the primary pointing sensor on two balloon-borne telescopes. In Fall, 2004, it flew on the InFOCμs payload from Ft. Sumner, NM[4]. The star camera performed to the required specification, providing a stable attitude during the entire 20 hour flight. It flew again the following year on the BLAST four day flight from Kiruna, Sweden. Because of its importance to the success of the mission, the OSC was constructed and flown for redundancy.

The second star camera enables us to continuously measure the precision of the star camera pointing solution. A comparison of many simultaneous solutions from both cameras is shown in Figure 5. The angular offset between the two star cameras was measured from simultaneous pointing solutions while the telescope was stationary on the ground.

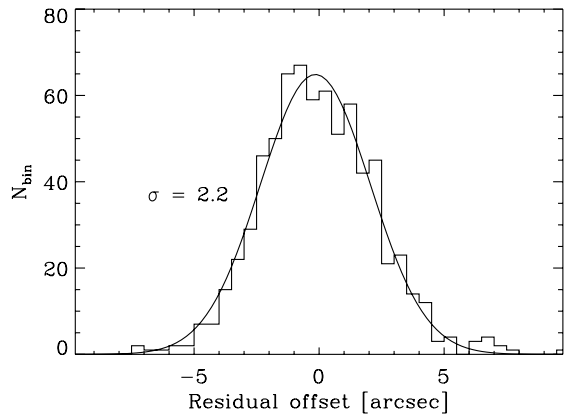

Fig. 5: The angular offset between the two star cameras was measured from 721 pointing solutions from simultaneous images over 28 minutes while the telescope was stationary on the ground. The mean offset was removed, and the residual offset is shown here as a histogram. The standard deviation of the Gaussian fit to the histogram is 2.2″. Assuming the errors in the two star camera pointing solutions are independent, the absolute uncertainty in an individual pointing solution inferred for one star camera is 2.2″ / √2 = 1.6″ rms.

A Gaussian fit to the offsets indicates the absolute uncertainty in an individual pointing solution for one star camera is < 2″ rms.

Both systems performed reliably throughout the four day BLAST flight. A sample 60 seconds of flight data from the ISC are shown in Figure 6. These data were taken during a scan of the Cygnus X region of the sky. The camera was set for an integration time of 30 ms while the gondola slewed in azimuth. The trigger was timed to allow a 200 ms integration time during the turnaround. With a 30 ms integration time the ISC was able to acquire at least 2 stars, and frequently > 3, enabling a pointing solution. With a 200 ms integration time the ISC was able to acquire > 8 stars in each exposure.

Pointing errors occasionally resulted when the gondola was in motion and the exposure time was set too high. In these instances the star camera would match the streaked star positions to an incorrect star field, potentially confusing the pointing solution. However, the main attitude determination software uses the other pointing sensors to determine the validity of the pointing solution returned by the star camera. The data from these other sensors are used to restore the star camera solution.

After the flight the star camera and gyroscope pointing solutions were combined to reconstruct the telescope attitude. Images with valid star camera solutions can be used to identify stars in sequential images that have too few sources to find a solution in flight. The gyroscope information can also be used to interpolate between star camera pointing solutions. Post flight analysis shows the star cameras are capable of providing a reconstructed pointing solution with an absolute accuracy < 5″.

## CONCLUSIONS

The BLAST star camera has been shown in two flights to provide reliable attitude information for balloon-borne payloads. In typical daytime float conditions the ISC can provide real-time pointing solutions with an absolute accuracy < 5″, at a frequency of 1 Hz. The instrument can run completely autonomously, or can be triggered externally to permit controlled synchronization. Sophisticated software analyzes the star fields, regulates the system housekeeping, and outputs a video signal displaying the results. This signal is transmitted to the ground station for visual confirmation of the instrument status.

Software is currently being developed that will enable the star camera to find a "lost in space" pointing solution. This means the star camera will not need a coarse initial guess to find its pointing. This fix would enable the star camera to run independently of any coarse sensors, and recover independently from the type of erroneous solutions seen in the 2005 flight.

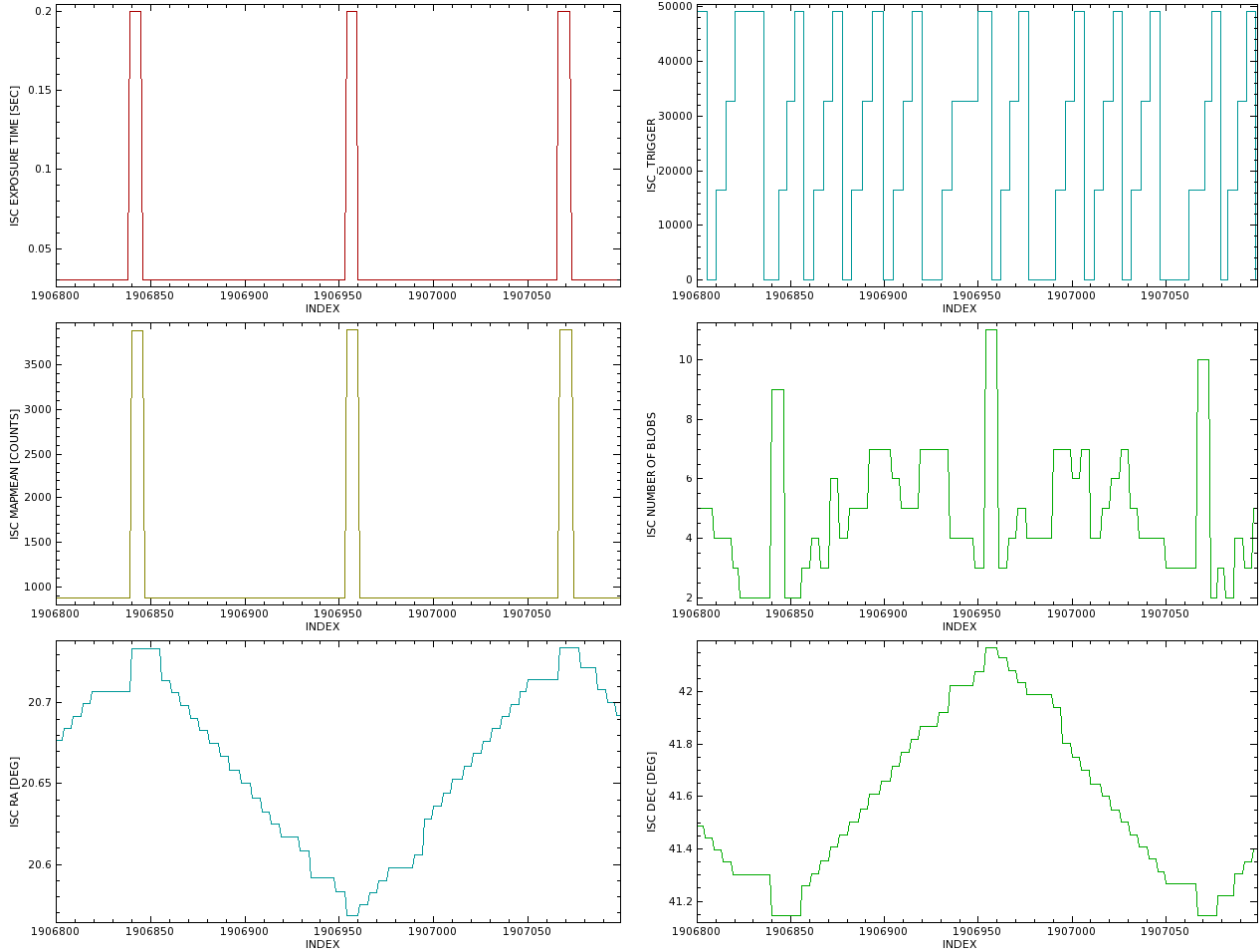

Fig. 6: 60 seconds of flight data were taken during a scan of the Cygnus X region of the sky. The camera was set for an integration time of 30 ms while the gondola slewed in azimuth. The trigger was timed to allow a 200 ms integration time during the turnaround. The number of stars being used in a pointing solution is indicated in the "ISC Number of Blobs" graph. With a 30 ms integration time ISC was able to acquire at least 2 stars, and frequently > 3, enabling a pointing solution. With a 200 ms integration time the ISC was able to acquire > 8 stars in each exposure.

## ACKNOWLEDGEMENTS

This work was funded under NASA grant NAG5-12785.

We gratefully acknowledge the efforts of the entire BLAST collaboration for assisting with the testing, alignment and operation of these instruments in the field and in flight. We thank the InFOCμs team from NASA Goddard Space Flight Center and Nagoya University (Japan) for providing field support and a balloon-borne test bed for the star camera. All machining work was preformed by the machine shop at the University of Pennsylvania. Adria Roode ran Modtran simulations to predict daytime sky brightness at float. The star camera design was loosely based on work done by William Craig's group at Lawrence Livermore National Labs.